\documentclass[a4paper,fleqn,usenatbib]{mnras}

\usepackage[T1]{fontenc}
\usepackage{ae,aecompl}

\usepackage{graphicx}	
\usepackage{amsmath}	
\usepackage{amssymb}	

\title[Precise CCD positions of Triton]{Precise CCD positions of Triton in 2014-2016 using the newest Gaia DR1 star catalogue}

\author[N. Wang et al.]{
N. Wang,$^{1,2}$
Q. Y. Peng,$^{1,2}$\thanks{E-mail: tpengqy@jnu.edu.cn (KTS)}
H. W. Peng,$^{1,2,3,4}$
H. J. Xie,$^{1,2}$
S. Ma,$^{1,2}$
Q. F. Zhang$^{1,2}$
\\
$^{1}$Department of Computer Science, Jinan University, Guangzhou 510632, China\\
$^{2}$Sino-French Joint Laboratory for Astrometry, Dynamics and Space Science, Jinan University, Guangzhou 510632, China\\
$^{3}$Yunnan Observatories, Chinese Academy of Sciences, Kunming 650216, China\\
$^{4}$University of Chinese Academy of Sciences, Beijing 100049, China
}

\date{Accepted 2017 March 1. Received 2017 February 26; in original form 2017 January 24}

\pubyear{2015}

\begin{document}
\label{firstpage}
\pagerange{\pageref{firstpage}--\pageref{lastpage}}
\maketitle

\begin{abstract}
755 new CCD observations during the years 2014-2016 have been reduced to derive the precise positions of Triton, the first satellite of Neptune. The observations were made by the 1 m telescope at Yunnan Observatory over fifteen nights. The positions of Triton are measured with respect to the stars in Gaia DR1 star catalogue. The theoretical position of Triton was retrieved from the Jet Propulsion Laboratory (JPL) ephemeris nep081xl, while the position of Neptune was obtained from the JPL ephemeris DE431mx. Our results show that the mean O-Cs (observed minus computed) are 0.042 and -0.006 arcsec in right ascension and declination respectively. The dispersions of our observations are estimated at about 0.012 arcsec in each direction.
\end{abstract}

\begin{keywords}
methods: observational -- techniques: image processing -- astrometry -- planets and satellites: general.
\end{keywords}



\section{Introduction}
Triton, the largest moon of Neptune, was discovered on October 10, 1846. It has an retrograde, inclined and circular orbit. This unusual configuration has led to the belief that Triton originally orbited the Sun before being captured in orbit around Neptune \citep{McCord,McKin}. Therefore Triton hold important clues to the evolution of the solar system. In recent years, generous observations of Triton have been made \citep{Veiga+1996,Veiga+etal+1998,Stone+etal+2000,Stone+etal+2001,Vieira+04,Qiao+etal+2007, Qiao+etal+2014} and the precision of these observations is usually about 0.05 to 0.5 arcsec \citep{Emelyanov+etal+2015}. As the newest Gaia DR1 star catalogue is available {\citep{gaia+16a,gaia+16b}, the precision of Triton must be further improved and a quite great CCD field of view allows us to calibrate its geometry accurately \citep{Peng+12}. Furthermore, the high precision of Triton can help improve the precision of Neptune. At present even with the modern planetary ephemeris DE431, measurement uncertainties such as the Earth's atmosphere and star catalog uncertainties limit position accuracies of Neptune to several thousand kilometers \citep{Willim+etal+2014}. From the statistics conducted by Folkner et al. \citep{Willim+etal+2014} in their figure 67-76, most astrometric observations of Neptune are not more accurate than 0.2 arcsec except the observations made by Voyager 2. As the position and motion of the planets can be deduced from the ephemerides of the satellites \citep{Robert+etal+2011}, the high precision of Triton can help improve the precision of Neptune. Besides, the continuation of observing campaign is necessary for the analysis of long-term physical element, such as the accuracy of the dynamical models \citep{Arlot+etal+2012}. Thus we must never stop observing, since an observation not made is a missing observation.

The contents of this paper are arranged as follows. In Section 2, the observations are described. Section 3 presents the method of reduction. In Section 4, we make comparisons and analysis of the residuals. Finally, in Section 5, we draw a conclusion.

\section{Observations}
\label{sect:Obs}
Since 2014, we have been engaged in a systematic observation of Triton. All the observations were made with the 1 m telescope at Yunnan Observatory. The site (i.e. IAU code 286) is at longitude E100${\degr}$ 01${\arcmin}$ 51${\arcsec}$, latitude N26${\degr}$ 42${\arcmin}$ 32${\arcsec}$ and height 2000 m above sea level. A total of 755 frames of CCD images have been obtained for the satellite as well as 398 frames of CCD calibration images. The exposure time for each CCD frame ranged from 30 to 120 s, depending on the meteorological conditions. For more instrumental details for the reflector and the CCD detector, see Table 1. In Table 2, the observational sets for Triton and calibration fields are given for each night. Due to the poor weather conditions or the limited time-allocation on October 16, 2014, October 18, 2014, November 3, 2015 and November 21-23, 2016, no calibration fields could be used to derive the geometric distortion (called GD hereafter) patterns of the field of view. Then the GD pattern of the nearest date is applied for as it was done in our previous work \citep{Peng+15,Wang+etal+2015}.
\begin{table}
\caption{\label{tab:Table1}Specifications of the 1 m telescope and CCD detector} \centering
\begin{tabular}{rr}
\hline
Approximate focal length             &1330cm\\
\hline
F-Ratio                           &13\\
Diameter of primary mirror        &100cm\\
CCD field of view            &${7.1\arcmin}$${\times}$${7.1\arcmin}$\\
Size of CCD array            &2048${\times}$2048\\
Size of pixel               &${13.5\mu}{\times}{13.5\mu}$\\
Approximate angular extent per pixel  &0${\farcs}$21/pixel\\
\hline
\end{tabular}
\end{table}
\begin{table}
\caption{\label{tab:Table2}Observations for Triton and calibration fields. Column 3 and Column 4 list the number for a dense star field and Triton, respectively. Column 5 lists the filter.}
\centering
\begin{tabular}{ccccc}
\hline
Obs Date     &Calibration fields& &Triton&filter\\
        &Dense Star Fields&No. &No. &\\
\hline
2014-10-16&No data available&    &28      &N\\
2014-10-17&NGC7092          &29  &41      &N\\
2014-10-18&No data available&    &37      &N\\
2015-11-03&No data available&    &21      &I\\
2015-11-04&NGC7092          &31  &20      &I\\
2016-09-25&NGC1664          &49  &42      &I\\
2016-09-26&NGC1664          &49  &42      &I\\
2016-09-27&NGC7209          &46  &5       &I\\
2016-10-22&NGC7209          &48  &71      &I\\
2016-10-23&NGC7209          &48  &70      &I\\
2016-10-24&NGC7209          &49  &69      &I\\
2016-11-20&NGC1664          &49  &68      &I\\
2016-11-21&No data available&    &80      &I\\
2016-11-22&No data available&    &86      &I\\
2016-11-23&No data available&    &75      &I\\
     total&       &398   &755\\
\hline
\end{tabular}
\end{table}

\section{Astrometric Reduction}
\label{sect:Red}
The reduction procedures were carried out according to \citet{Peng+12} and involved the following: (1) The calibration fields and Triton were processed including determining the centers of stars and matching the stars in each CCD frames with those in some reference star catalogue. Here, the Gaia DR1 star catalogue was chosen as the reference star catalogue because of the large number of reference stars available. However, due to the shortage of proper motions for most stars in Gaia DR1 star catalogue, a value of 0 was adopted for the value of proper motion. As the positions of stars in Gaia DR1 are obtained at epoch J2015.0 \citep{Linde+16} and our observations are made near the epoch of Gaia DR1, the effect of proper motions for faint stars can be neglected. Furthermore, as the locations of the stars are randomly distributed in the different fields of view when Triton is observed, the average effect of proper motions to the positional reduction of Triton can be quite cancelled out; (2) An accurate reduction was done including the computations of topocentric apparent position and atmospheric refraction for each matched star in each CCD frame. Then the standard coordinates of each star were calculated. During the computation of the theoretical positions of Triton, the modern ephemeris nep081xl/DE431mx developed by JPL via their Web site http://ssd.jpl.nasa.gov/ is used; (3) The GD patterns were derived from the calibration fields and the pixel positions of stars and Triton were both corrected by using the derived GD. Then the observed position of Triton was obtained by the transformation of the four parameters.

The (O-C) residuals larger than 3 times standard deviation have been rejected and 755 (O-C) residuals were obtained. Figure~\ref{fig:figure1} shows the (O-C) residuals of Triton with respect to the observational epochs. In Figure~\ref{fig:figure1} the results with and without GD corrections are also displayed. Table~\ref{tab:Table3} shows the statistics of (O-C) residuals of Triton before and after GD correction. It can be seen the precision is improved after GD correction.

\begin{figure*}
	\includegraphics[width=14cm]{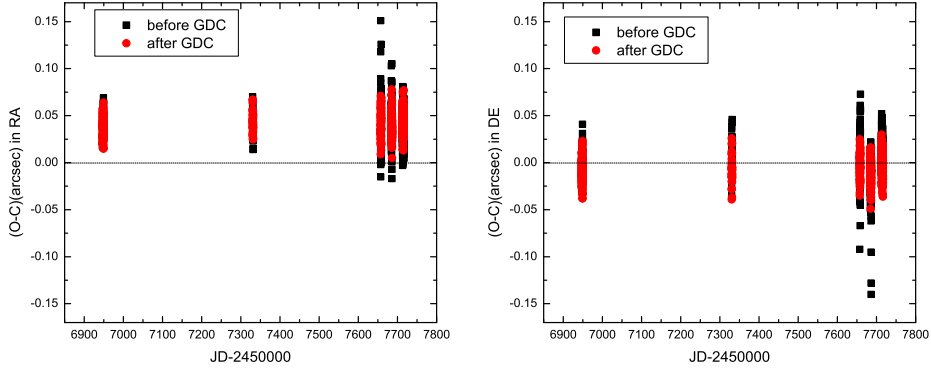}
    \caption{(O-C) residuals of the positions of Triton in three years. The dark points represent the (O-C) before GD correction and the red ones represent the (O-C) after GD correction. The ephemerides used are JPL nep081xl/DE431mx and the reference star catalogue is Gaia DR1 star catalogue.}
    \label{fig:figure1}
\end{figure*}

\begin{table}
\caption{\label{tab:Table3}Statistics of (O-C) residuals for Triton before and after GD correction for Each Data Set. Columns 1 is the date. The second column ("Before GDC" or "After GDC") shows the statistics in both directions before or after GD correction. The following columns list the mean (O-C) and its standard deviation (SD) in right ascension and declination, respectively. All units are in arcseconds. The ephemeris used is JPL nep081xl/DE431mx and the reference star catalogue is Gaia DR1 star catalogue.} \centering
\begin{tabular}{{@{}ccccrc@{}}}
\hline  Date & &$<$O-C$>$&SD&$<$O-C$>$&SD\\
     &&\multicolumn{2}{c}{RA}     &\multicolumn{2}{c}{DEC}\\
\hline
2014 &Before GDC  &0.039  &0.011  &-0.005   &0.014\\
     &After GDC   &0.037  &0.011  &-0.008  &0.013\\
2015 &Before GDC  &0.040  &0.013  &0.013   &0.019\\
     &After GDC   &0.044  &0.011  &-0.006  &0.015\\
2016 &Before GDC  &0.039  &0.019  &0.004   &0.022\\
     &After GDC   &0.043  &0.011  &-0.006  &0.012\\
total &Before GDC  &0.039  &0.018  &0.003   &0.022\\
      &After GDC   &0.042  &0.012  &-0.006  &0.012\\
 \hline
\end{tabular}
\end{table}

\section{Discussions}
\label{sect:diss}
In order to make a comparison, the stars in each CCD frame are also matched with the stars from USNO CCD Astrograph Catalogue 4 (UCAC4; \citet{Zacharias+etal+2013}) star catalogue. Figure~\ref{fig:figure2} shows the (O-C) residuals by using both Gaia DR1 and UCAC4 star catalogues and Table~\ref{tab:Table4} shows the statistics of (O-C) residuals of Triton by using the two star catalogues. It appears obviously that by using the Gaia DR1 star catalogue the solution has much better agreements and much smaller dispersions in each direction. The mean (O-C) is better than 0.05 arcsec in each direction and the dispersion is better than 0.02 arcsec in each direction. Figure~\ref{fig:figure3} shows a scatter plot of the (O-C) residuals in right ascension and declination with respect to the two different catalogues. It appears that much smaller dispersions can be obtained when Gaia DR1 star catalogue is referred to.

\begin{figure*}
	\includegraphics[width=14cm]{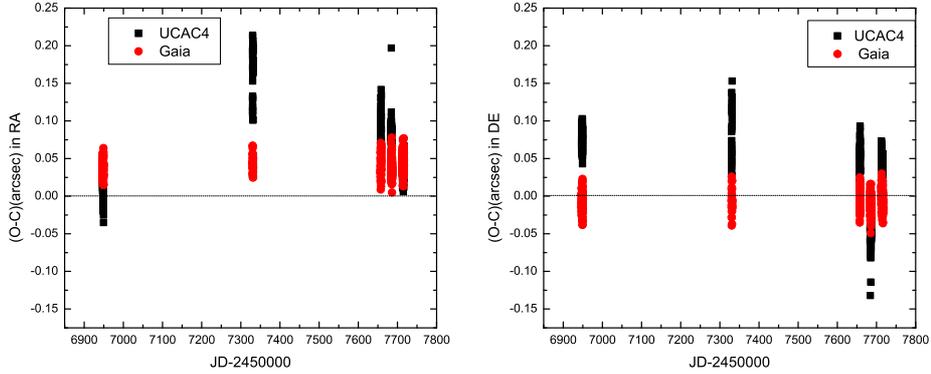}
    \caption{(O-C) residuals of the positions of Triton during 2014-2016 by using different star catalogues. The dark points represent the (O-C)s using UCAC4 star catalogue and the red ones represent the (O-C)s using Gaia DR1 star catalogue.}
    \label{fig:figure2}
\end{figure*}

\begin{table}
\caption{\label{tab:Table4}Statistics of (O-C) residuals for Triton using different star catalogue. Columns 1 is the date. The second column shows the star catalogues. The following columns list the mean (O-C) and its standard deviation (SD) in right ascension and declination, respectively. All units are in arcseconds. The ephemeris used is JPL nep081xl/DE431mx.} \centering
\begin{tabular}{{@{}ccrrrr@{}}}
\hline  Date & &$<$O-C$>$&SD&$<$O-C$>$&SD\\
     &&\multicolumn{2}{c}{RA}     &\multicolumn{2}{c}{DEC}\\
\hline
2014-10-16 &UCAC4  &0.028  &0.008  &0.071   &0.009\\
           &Gaia   &0.036  &0.011  &-0.008  &0.010\\
2014-10-17 &UCAC4  &0.018  &0.006  &0.080   &0.010\\
           &Gaia   &0.037  &0.009  &-0.008  &0.011\\
2014-10-18 &UCAC4 &-0.007  &0.010  &0.074   &0.013\\
           &Gaia   &0.036  &0.013  &-0.009  &0.017\\
2015-11-03 &UCAC4  &0.157  &0.038  &0.060   &0.029\\
           &Gaia   &0.047  &0.012  &-0.005  &0.017\\
2015-11-04 &UCAC4  &0.173  &0.028  &0.116   &0.014\\
           &Gaia   &0.041  &0.009  &-0.007  &0.013\\
2016-09-25 &UCAC4  &0.073  &0.020  &0.041   &0.018\\
           &Gaia   &0.040  &0.014  &-0.005  &0.013\\
2016-09-26 &UCAC4  &0.094  &0.025  &0.053   &0.024\\
           &Gaia   &0.045  &0.014  &-0.002  &0.011\\
2016-09-27 &UCAC4  &0.093  &0.028  &0.009   &0.023\\
           &Gaia   &0.052  &0.020  &-0.004  &0.016\\
2016-10-22 &UCAC4  &0.088  &0.016  &-0.052  &0.013\\
           &Gaia   &0.040  &0.010  &-0.006  &0.010\\
2016-10-23 &UCAC4  &0.055  &0.010  &-0.067  &0.010\\
           &Gaia   &0.040  &0.010  &-0.009  &0.010\\
2016-10-24 &UCAC4  &0.062  &0.007  &-0.044  &0.013\\
           &Gaia   &0.034  &0.010  &-0.012  &0.010\\
2016-11-20 &UCAC4  &0.033  &0.009  &0.050   &0.011\\
           &Gaia   &0.048  &0.009  &0.002   &0.012\\
2016-11-21 &UCAC4  &0.047  &0.009  &0.034   &0.012\\
           &Gaia   &0.049  &0.010  &-0.003  &0.011\\
2016-11-22 &UCAC4  &0.041  &0.009  &0.030   &0.009\\
           &Gaia   &0.043  &0.010  &-0.007  &0.010\\
2016-11-23 &UCAC4  &0.041  &0.012  &0.029   &0.013\\
           &Gaia   &0.046  &0.010  &-0.010  &0.011\\
total      &UCAC4  &0.056  &0.038  &0.020   &0.052\\
           &Gaia   &0.042  &0.012  &-0.006  &0.012\\
 \hline
\end{tabular}
\end{table}

\begin{figure}
	\includegraphics[width=8cm]{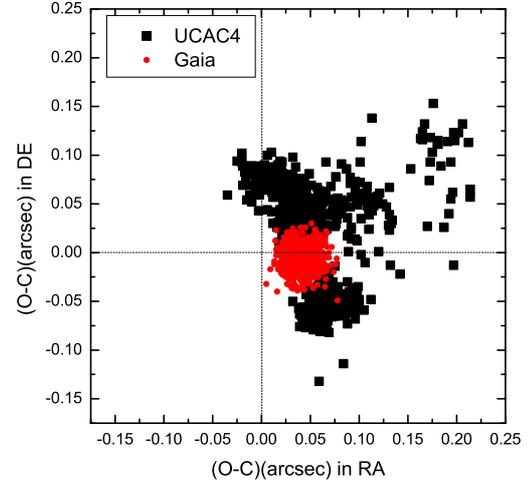}
    \caption{(O-C) residuals of the Triton during 2014-2016. The dark points represent the (O-C) residuals using UCAC4 star catalogue and the red ones represent the (O-C) residuals using Gaia DR1 star catalogue.}
    \label{fig:figure3}
\end{figure}

To analyse the observations, we also compared our observations with different planetary and satellite ephemerides which can be downloaded from IMCCE Web site http://www.imcce.fr/ and JPL Web site http://ssd.jpl.nasa.gov/. Table~\ref{tab:Table5} shows the (O-C) residuals by using different ephemerides. From Table~\ref{tab:Table5}, we can see that all ephemerides show a small dispersion, but the agreement varies greatly. Specifically, when the same planetary ephemeris DE431 is used, the solutions show a similar agreement. However, when the same satellite ephemeris developed by \citet{Emelyanov+etal+2015} is used, the solutions show a very different agreement. Thus the difference between the satellite ephemerides is very small, but the differences among the planetary ephemerides are evident. Obviously, DE431 appears to be a relatively better theory.

\begin{table}
\caption{\label{tab:Table5}Statistics of (O-C) residuals for Triton by using different satellite ephemerides and planetary ephemerides. Columns 1 is the number of observations, column 2 the planetary ephemeris and column 3 the satellite ephemerides. The following columns list the mean (O-C) and its standard deviation (SD) in right ascension and in declination, respectively. Emelyanov15 is a short form of Emelyanov et al. (2015). All units are in arcseconds.} \centering
\begin{tabular}{{@{}ccccrc@{}}}
\hline
  planetary &satellite &$<$O-C$>$&SD&$<$O-C$>$&SD\\
    ephemeris&ephemeris&\multicolumn{2}{c}{RA}     &\multicolumn{2}{c}{DEC}\\
\hline
INPOP08  &Emelyanov15&0.189 &0.016 & 0.022 &0.015\\
INPOP10  &Emelyanov15&0.083 &0.013 &-0.099 &0.013\\
INPOP13C &Emelyanov15&0.062 &0.013 &-0.059 &0.013\\
EPM2015  &Emelyanov15&0.062 &0.013 &-0.014 &0.013\\
DE405    &Emelyanov15&0.149 &0.016 &-0.028 &0.014\\
DE421    &Emelyanov15&0.057 &0.013 &-0.018 &0.013\\
DE431    &Emelyanov15&0.042 &0.013 &-0.007 &0.013\\
DE431    &JPL nep081xl&0.042 &0.012 &-0.006 &0.012\\
 \hline
\end{tabular}
\end{table}

To compare our observations with other ones, we also list some major observational statistics of Triton. Table~\ref{tab:Table6} lists some typical residuals of the observations. It can be seen that our observations have a smaller dispersion. In Table~\ref{tab:Table6}, the mean (O-C) and its standard deviation come from the literature of the author, the theoretical positions are obtained from JPL ephemeris.
\begin{table}
\caption[]{Compared with other observations. Column 1 lists the author, Column 2 the number of observations. The following columns list the mean (O-C) and its standard deviation (SD) in right ascension and in declination, respectively. All units are in arcseconds.
\label{tab:Table6}}
\setlength{\tabcolsep}{1pt}
\small
\centering
 \begin{tabular}{ccccccccccccc}
  \hline\noalign{\smallskip}
Author& No. &  $<$O-C$>$&SD&$<$O-C$>$&SD\\
     &&\multicolumn{2}{c}{RA}     &\multicolumn{2}{c}{DEC}\\
Veiga \& Vieira Martins(1996)    & 433        &-0.001&0.093 &0.032  &0.116\\
Veiga \& Vieira Martins(1998)    & 759        & 0.014&0.086 &-0.007 &0.082\\
Stone (2001)                     & 188        &-0.027&0.120 &-0.048 &0.113\\
Vieira Martins et al. (2004)     &  63        & 0.041&0.151 &0.060  &0.081\\
Qiao et al. (2007)               & 943        &0.062 &0.062 &-0.036 &0.053\\
Qiao et al. (2014)               &1095        &0.023 &0.062 &-0.029 &0.067\\
this work                        & 755        &0.042 &0.012 &-0.006 &0.012\\
  \noalign{\smallskip}\hline
\end{tabular}
\end{table}

Table~\ref{tab:Table7} lists an extract of our observed topocentric astrometric positions of Triton. The data are presented in the following form: JD denotes the exposure middle time of each CCD frame in the form of Julian Date (UTC). RA, expressed in hours, minutes and seconds are the positions of Triton in right ascension. DE, expressed in degrees, arc minutes and arc seconds are the positions of Triton in declination. The complete data can be obtained in the data base of Natural Satellites of the IMCCE.

\begin{table}
\caption{\label{tab:Table7} Extract of the observations. The first column lists the exposure middle time of each CCD frame in form of Julian Date (UTC). RA and DE are the observed topocentric astrometric positions in right ascension and declination, respectively. The whole table is available on the Web site http://lnfm1.sai.msu.ru/neb/nss/bnepomae.htm.}
\centering
\begin{tabular}{cccccc}
\hline  Date &RA &DE\\

JD&h m s& ${\degr}$ ${\arcmin}$ ${\arcsec}$\\
\hline
2456947.01326& 22 27 53.1995& -10 25 47.058\\
2456947.01664& 22 27 53.1882& -10 25 47.142\\
2456947.02484& 22 27 53.1623& -10 25 47.373\\
......&......&......\\
2457716.05200& 22 43 44.6339& -08 59 24.261\\
2457716.05281& 22 43 44.6352& -08 59 24.261\\
2457716.05362& 22 43 44.6361& -08 59 24.232\\
 \hline
\end{tabular}
\end{table}

\section{Conclusions}
\label{sect:con}
In this paper, we present 755 CCD observations of Triton, the first satellite of Neptune, taken with the 1 m telescope at Yunnan Observatory. During the reduction, the newest Gaia DR1 star catalogue are used to match the stars in the field of view. Comparisons have been made between the observed positions and the computed ones. The result has shown that after GD correction the mean (O-C) are 0.042 and -0.006 arcsec in right ascension and in declination with respect to JPL ephemeris, respectively. The dispersions of our observations are estimated at about 0.012 arcsec in each direction.

\section*{Acknowledgements}
We acknowledge the support of the staff of the 1 m telescope at Yunnan Observatory. This research work is financially supported by the National Natural Science Foundation of China (grant nos. U1431227, 11273014) and partly by the Fundamental Research Funds for the Central Universities. This work has made use of data from the European Space Agency (ESA) mission {\it Gaia} (\url{https://www.cosmos.esa.int/gaia}), processed by
the {\it Gaia} Data Processing and Analysis Consortium (DPAC, \url{https://www.cosmos.esa.int/web/gaia/dpac/consortium}). Funding for the DPAC has been provided by national institutions, in particular the institutions participating in the {\it Gaia} Multilateral Agreement.


\bibliographystyle{mnras}
\bibliography{bibtex} 

\begin{thebibliography}{}
\makeatletter
\relax
\def\mn@urlcharsother{\let\do\@makeother \do\$\do\&\do\#\do\^\do\_\do\%\do\~}
\def\mn@doi{\begingroup\mn@urlcharsother \@ifnextchar [ {\mn@doi@}
  {\mn@doi@[]}}
\def\mn@doi@[#1]#2{\def\@tempa{#1}\ifx\@tempa\@empty \href
  {http://dx.doi.org/#2} {doi:#2}\else \href {http://dx.doi.org/#2} {#1}\fi
  \endgroup}
\def\mn@eprint#1#2{\mn@eprint@#1:#2::\@nil}
\def\mn@eprint@arXiv#1{\href {http://arxiv.org/abs/#1} {{\tt arXiv:#1}}}
\def\mn@eprint@dblp#1{\href {http://dblp.uni-trier.de/rec/bibtex/#1.xml}
  {dblp:#1}}
\def\mn@eprint@#1:#2:#3:#4\@nil{\def\@tempa {#1}\def\@tempb {#2}\def\@tempc
  {#3}\ifx \@tempc \@empty \let \@tempc \@tempb \let \@tempb \@tempa \fi \ifx
  \@tempb \@empty \def\@tempb {arXiv}\fi \@ifundefined
  {mn@eprint@\@tempb}{\@tempb:\@tempc}{\expandafter \expandafter \csname
  mn@eprint@\@tempb\endcsname \expandafter{\@tempc}}}

\bibitem[\protect\citeauthoryear{{Arlot}, {Desmars}, {Lainey}  \&
  {Robert}}{{Arlot} et~al.}{2012}]{Arlot+etal+2012}
{Arlot} J.-E.,  {Desmars} J.,  {Lainey} V.,   {Robert} V.,  2012, \mn@doi
  [\planss] {10.1016/j.pss.2012.10.002}, \href
  {http://adsabs.harvard.edu/abs/2012P%26SS...73...66A} {73, 66}

\bibitem[\protect\citeauthoryear{{Emelyanov} \& {Samorodov}}{{Emelyanov} \&
  {Samorodov}}{2015}]{Emelyanov+etal+2015}
{Emelyanov} N.~V.,  {Samorodov} M.~Y.,  2015, \mn@doi [\mnras]
  {10.1093/mnras/stv2116}, \href
  {http://adsabs.harvard.edu/abs/2015MNRAS.454.2205E} {454, 2205}

\bibitem[\protect\citeauthoryear{{Folkner}, {Williams}, {Boggs}, {Park}  \&
  {Kuchynka}}{{Folkner} et~al.}{2014}]{Willim+etal+2014}
{Folkner} W.~M.,  {Williams} J.~G.,  {Boggs} D.~H.,  {Park} R.~S.,   {Kuchynka}
  P.,  2014, Interplanetary Network Progress Report, \href
  {http://adsabs.harvard.edu/abs/2014IPNPR.196C...1F} {196, 1}

\bibitem[\protect\citeauthoryear{{Gaia Collaboration} et~al.,}{{Gaia
  Collaboration} et~al.}{2016a}]{gaia+16a}
{Gaia Collaboration} et~al., 2016a, \mn@doi [\aap]
  {10.1051/0004-6361/201629272}, \href
  {http://adsabs.harvard.edu/abs/2016A%26A...595A...1G} {595, A1}

\bibitem[\protect\citeauthoryear{{Gaia Collaboration} et~al.,}{{Gaia
  Collaboration} et~al.}{2016b}]{gaia+16b}
{Gaia Collaboration} et~al., 2016b, \mn@doi [\aap]
  {10.1051/0004-6361/201629512}, \href
  {http://adsabs.harvard.edu/abs/2016A%26A...595A...2G} {595, A2}

\bibitem[\protect\citeauthoryear{{Lindegren} et~al.,}{{Lindegren}
  et~al.}{2016}]{Linde+16}
{Lindegren} L.,  et~al., 2016, \mn@doi [\aap] {10.1051/0004-6361/201628714},
  \href {http://adsabs.harvard.edu/abs/2016A%26A...595A...4L} {595, A4}

\bibitem[\protect\citeauthoryear{{McCord}}{{McCord}}{1966}]{McCord}
{McCord} T.~B.,  1966, \mn@doi [\aj] {10.1086/109967}, \href
  {http://adsabs.harvard.edu/abs/1966AJ.....71..585M} {71, 585}

\bibitem[\protect\citeauthoryear{{McKinnon}}{{McKinnon}}{1984}]{McKin}
{McKinnon} W.~B.,  1984, \mn@doi [\nat] {10.1038/311355a0}, \href
  {http://adsabs.harvard.edu/abs/1984Natur.311..355M} {311, 355}

\bibitem[\protect\citeauthoryear{{Peng}, {Vienne}, {Zhang}, {Desmars}, {Yang}
  \& {He}}{{Peng} et~al.}{2012}]{Peng+12}
{Peng} Q.~Y.,  {Vienne} A.,  {Zhang} Q.~F.,  {Desmars} J.,  {Yang} C.~Y.,
  {He} H.~F.,  2012, \mn@doi [\aj] {10.1088/0004-6256/144/6/170}, \href
  {http://adsabs.harvard.edu/abs/2012AJ....144..170P} {144, 170}

\bibitem[\protect\citeauthoryear{{Peng}, {Wang}, {Vienne}, {Zhang}, {Li}  \&
  {Meng}}{{Peng} et~al.}{2015}]{Peng+15}
{Peng} Q.~Y.,  {Wang} N.,  {Vienne} A.,  {Zhang} Q.~F.,  {Li} Z.,   {Meng}
  X.~H.,  2015, \mn@doi [\mnras] {10.1093/mnras/stv469}, \href
  {http://adsabs.harvard.edu/abs/2015MNRAS.449.2638P} {449, 2638}

\bibitem[\protect\citeauthoryear{{Qiao} et~al.,}{{Qiao}
  et~al.}{2007}]{Qiao+etal+2007}
{Qiao} R.~C.,  et~al., 2007, \mn@doi [\mnras]
  {10.1111/j.1365-2966.2007.11526.x}, 376, 1707

\bibitem[\protect\citeauthoryear{{Qiao} et~al.,}{{Qiao}
  et~al.}{2014}]{Qiao+etal+2014}
{Qiao} R.~C.,  et~al., 2014, \mn@doi [\mnras] {10.1093/mnras/stu566}, \href
  {http://adsabs.harvard.edu/abs/2014MNRAS.440.3749Q} {440, 3749}

\bibitem[\protect\citeauthoryear{{Robert} et~al.,}{{Robert}
  et~al.}{2011}]{Robert+etal+2011}
{Robert} V.,  et~al., 2011, \mn@doi [\mnras]
  {10.1111/j.1365-2966.2011.18747.x}, \href
  {http://adsabs.harvard.edu/abs/2011MNRAS.415..701R} {415, 701}

\bibitem[\protect\citeauthoryear{{Stone}}{{Stone}}{2000}]{Stone+etal+2000}
{Stone} R.~C.,  2000, \mn@doi [\aj] {10.1086/301577}, \href
  {http://adsabs.harvard.edu/abs/2000AJ....120.2124S} {120, 2124}

\bibitem[\protect\citeauthoryear{{Stone}}{{Stone}}{2001}]{Stone+etal+2001}
{Stone} R.~C.,  2001, \mn@doi [\aj] {10.1086/323549}, \href
  {http://adsabs.harvard.edu/abs/2001AJ....122.2723S} {122, 2723}

\bibitem[\protect\citeauthoryear{{Veiga} \& {Martins}}{{Veiga} \&
  {Martins}}{1998}]{Veiga+etal+1998}
{Veiga} C.~H.,  {Martins} R.~V.,  1998, \mn@doi [\aaps] {10.1051/aas:1998268},
  \href {http://adsabs.harvard.edu/abs/1998A%26AS..131..291V} {131, 291}

\bibitem[\protect\citeauthoryear{{Veiga} \& {Vieira Martins}}{{Veiga} \&
  {Vieira Martins}}{1996}]{Veiga+1996}
{Veiga} C.~H.,  {Vieira Martins} R.,  1996, \aaps, \href
  {http://adsabs.harvard.edu/abs/1996A%26AS..120..107V} {120, 107}

\bibitem[\protect\citeauthoryear{{Vieira Martins}, {Veiga}, {Bourget}, {Andrei}
   \& {Descamps}}{{Vieira Martins} et~al.}{2004}]{Vieira+04}
{Vieira Martins} R.,  {Veiga} C.~H.,  {Bourget} P.,  {Andrei} A.~H.,
  {Descamps} P.,  2004, \mn@doi [\aap] {10.1051/0004-6361:200400024}, \href
  {http://adsabs.harvard.edu/abs/2004A%26A...425.1107V} {425, 1107}

\bibitem[\protect\citeauthoryear{{Wang}, {Peng}, {Zhang}, {Zhang}, {Li}  \&
  {Meng}}{{Wang} et~al.}{2015}]{Wang+etal+2015}
{Wang} N.,  {Peng} Q.~Y.,  {Zhang} X.~L.,  {Zhang} Q.~F.,  {Li} Z.,   {Meng}
  X.~H.,  2015, \mn@doi [\mnras] {10.1093/mnras/stv2236}, \href
  {http://adsabs.harvard.edu/abs/2015MNRAS.454.3805W} {454, 3805}

\bibitem[\protect\citeauthoryear{{Zacharias}, {Finch}, {Girard}, {Henden},
  {Bartlett}, {Monet}  \& {Zacharias}}{{Zacharias}
  et~al.}{2013}]{Zacharias+etal+2013}
{Zacharias} N.,  {Finch} C.~T.,  {Girard} T.~M.,  {Henden} A.,  {Bartlett}
  J.~L.,  {Monet} D.~G.,   {Zacharias} M.~I.,  2013, \mn@doi [\aj]
  {10.1088/0004-6256/145/2/44}, \href
  {http://adsabs.harvard.edu/abs/2013AJ....145...44Z} {145, 44}

\makeatother
\end{thebibliography}

\bsp	
\label{lastpage}
\end{document}